\documentclass[aps,preprint,epsfig,rotate]{revtex4}
\usepackage{graphicx}
\usepackage{bm}
\usepackage{epsfig}
\input epsf
\begin{document}
%\begin{doublespace}
\title{Compact and accurate variational wave functions of three-electron
       atomic systems constructed from semi-exponential radial basis
       functions}

 \author{Alexei M. Frolov}
 \email[E--mail address: ]{afrolov@uwo.ca}

\affiliation{Department of Chemistry\\
 University of Western Ontario, London, Ontario N6H 5B7, Canada}

\date{\today}

\begin{abstract}

The semi-exponential basis set of radial functions (A.M. Frolov, Physics
Letters A {\bf 374}, 2361 (2010)) is used for variational computations of
bound states in three-electron atomic systems. It appears that
semi-exponential basis set has a substantially greater potential for
accurate variational computations of bound states in three-electron atomic
systems than it was originally anticipated. In particular, the 40-term
Larson's wave function improved with the use of semi-exponential radial
basis functions now produces the total energy \linebreak -7.4780581457
$a.u.$ for the ground $1^2S-$state in the ${}^{\infty}$Li atom (only one
spin function $\chi_1 = \alpha \beta \alpha - \beta \alpha \alpha$ was used
in these calculations). This variational energy is very close to the exact
ground state energy of the ${}^{\infty}$Li atom and it substantially lower
than the total energy obtained with the original Larson's 40-term wave
function (-7.477944869 $a.u.$).

PACS number(s): 03.65.Ge, 31.15.ac and 31.15.xf
\end{abstract}
\maketitle
%\noindent \vspace{0.2in}

\section{Introduction}\label{intro}

In this study we perform variational calculations of bound states in
three-electron atomic systems. A basis set of semi-exponential radial
functions \cite{Fro2010} is extensively used in our calculations. The main
goal is to solve the non-relativistic Schr\"{o}dinger equation $H \Psi = E
\Psi$, where $E < 0$ and bound state wave function $\Psi$ has the unit norm.
The general non-relativistic Hamiltonian $H$ of the three-electron atomic
problem is (see, e.g., \cite{LLQ})
\begin{eqnarray}
 H = -\frac{\hbar^2}{2 m_e} \Bigl[\nabla^2_1 + \nabla^2_2 +
      \nabla^2_3 + \frac{m_e}{M} \nabla^2_4 \Bigr]
     - \frac{Q e^2}{r_{14}} - \frac{Q e^2}{r_{24}}
     - \frac{Q e^2}{r_{34}} + \frac{e^2}{r_{12}} + \frac{e^2}{r_{13}}
     + \frac{e^2}{r_{23}} \label{Hamil}
\end{eqnarray}
where $\hbar = \frac{h}{2 \pi}$ is the reduced Planck constant, $m_e$ is the
electron mass and $e$ is the electric charge of electron. In this equation
and everywhere below in this study the subscripts 1, 2, 3 designate the
three electrons $e^-$, while the subscript 4 denotes the heavy nucleus with
the mass $M$ ($M \gg m_e$) and positive electric (nuclear) charge $Q e$. The
notations $r_{ij} = \mid {\bf r}_i - {\bf r}_j \mid = r_{ji}$ stand for the
six interparticle distances (= relative coordinates) defined in an arbitrary
four-body system and ${\bf r}_i$ ($i$ = 1, 2, 3, 4) are the Cartesian
coordinates of the four point particles. In Eq.(\ref{Hamil}) and everywhere
below in this work we shall assume that $(ij)$ = $(ji)$ = (12), (13), (14),
(23), (24), (34). Below only atomic units $\hbar = 1, \mid e \mid = 1, m_e =
1$ are employed. In these units the explicit form of the Hamiltonian $H$,
Eq.(\ref{Hamil}), is significantly simplified.

The main attention in this work is focused on numerical calculations of the
ground (doublet) $1^2S(L = 0)-$state (or $1^2S_{\frac12}(L = 0)-$state) of
the three-electron Li atom with the infinitely heavy nucleus, i.e. the
${}^{\infty}$Li atom. Considerations of other three-electron atoms, ions
and various positron containing atomic systems (e.g., HPs) can be performed
absolutely analogously (for more detail, see \cite{Fro2010}) and, therefore,
these systems will not be considered here.

The problem of highly accurate calculations of the bound states in
three-electron atomic systems has attracted continuing attention. The first
calculations of the Li atom with the truly correlated wave functions were
performed in 1936 \cite{JC}. A brief reviews of such calculations can be
found in \cite{McW1} (earlier works) and \cite{King1} (references up to 1997
are mentioned). The current bibliography on this subject includes almost one
thousand references and is increasing rapidly. The classical Hylleraas
method (see, e.g., \cite{King1}) is the method capable of predicting the
most highly accurate wave functions for bound states of three-electron
atomic systems. In this method to produce the most highly accurate wave
functions (e.g., for the ground (doublet) $1^2S$-state in the Li atom)
one needs to use many thousands of Hylleraas basis functions. The use of
extremely large basis sets is very inconvenient in many actual cases, since
it produces a number of computational problems. It is clear that the
classical Hylleraas method cannot be used to construct both compact and
accurate variational wave functions for three-electron systems. Indeed, it
contains essentially no control parameters which can be optimized by
increasing the overall efficiency of the method.

An alternative approach to variational bound state calculations in
three-electron atomic systems was proposed in \cite{Fro2010}. This approach
is based on the use of semi-exponential variational wave functions
\cite{Fro2010} and it allows one to construct very compact and accurate
variational wave functions for arbitrary three-electron atomic system. Each
of the semi-exponential basis functions depend upon all six interparticle
coordinates $r_{12}, r_{13}, \ldots, r_{34}$ \cite{Fro2010}. A high
efficiency of this new approach in actual applications and its superiority
over the classical Hylleraas expansion was demonstrated in \cite{Fro2010}.
As follows from the results of this study the semi-exponential basis set has
a substantially greater potential for highly accurate variational
computations of bound states in three-electron atomic systems than it was
anticipated earlier \cite{Fro2010}.

\section{Variational wave function}

The variational wave function of the doublet $S(L = 0)-$states of the
three-electron Li atom is written in the following general form
\begin{eqnarray}
 \Psi_{L=0} = \psi_{L=0}(A; \bigl\{ r_{ij} \bigr\}) (\alpha \beta \alpha
 - \beta \alpha \alpha) + \phi_{L=0}(B; \bigl\{ r_{ij} \bigr\}) (2 \alpha
 \alpha \beta  - \beta \alpha \alpha - \alpha \beta \alpha) \label{psi}
\end{eqnarray}
where $\psi_{L=0}(A; \bigl\{ r_{ij} \bigr\})$ and $\phi_{L=0}(B; \bigl\{
r_{ij} \bigr\})$ are the two independent spatial parts (= radial parts) of
the total wave function. Each of these two radial functions is, in fact, a
radial factor (for states with $L = 0$) in front of the corresponding
three-electron spin functions $\chi_1 = \alpha \beta \alpha - \beta \alpha
\alpha$ and $\chi_2 = 2 \alpha \alpha \beta  - \beta \alpha \alpha - \alpha
\beta \alpha$. Here the notations $\alpha$ and $\beta$ are the one-electron
spin-up and spin-down functions, respectively (see, e.g., \cite{Dir}). The
notations $A$ and $B$ in Eq.(\ref{psi}) mean that the two sets of non-linear
parameters associated with radial functions $\psi$ and $\phi$ can be
optimized independently. In the general case, each of the radial basis
functions explicitly depends upon all six interparticle (relative)
coordinates $r_{12}, r_{13}, r_{23}, r_{14}, r_{24}, r_{34}$. It is clear
that in actual variational calculations only one spin function, e.g., the
$\chi_1$ function, need be used.  Another useful trick (so-called
`doubling') is based on the use of the same set of non-linear parameters in
the two radial parts in Eq.(\ref{psi}).

In our earlier work \cite{Fro2010} we have introduced an advanced set of
radial basis functions for three-electron atomic calculations. In
\cite{Fro2010} this set was called the semi-exponential basis set. In
general, the semi-exponential variational expansion of the radial function
$\psi_{L=0}(A; \bigl\{ r_{ij} \bigr\})$ is written in the form
\begin{eqnarray}
 \psi_{L=0}(A; \bigl\{ r_{ij} \bigr\}) = \sum^N_{k=1} C_k r^{n_1(k)}_{23}
 r^{n_2(k)}_{13} r^{n_3(k)}_{12} r^{m_1(k)}_{14} r^{m_2(k)}_{24}
 r^{m_3(k)}_{34} exp(-\alpha_{k} r_{14} -\beta_{k} r_{24} -\gamma_{k}
 r_{34}) \label{semexp}
\end{eqnarray}
where $\alpha_k, \beta_k, \gamma_k$ ($k = 1, 2, \ldots, N$) are the varied
non-linear parameters. The presence of the varied non-linear parameters in
Eq.(\ref{semexp}) is the main and very important difference with the
traditional Hylleraas variational expansion (see, e.g., \cite{Lars}) for
which in Eq.(\ref{semexp}) we always have $\alpha_1 = \ldots = \alpha_N,
\beta_1 = \ldots = \beta_N$ and $\gamma_1 = \ldots = \gamma_N$. Note that
all matrix elements of the Hamiltonian, Eq.(\ref{Hamil}), and overlap matrix
needed in computations with the use of the semi-exponential basis,
Eq.(\ref{semexp}), contain the same three-electron integrals which arise for
the usual Hylleraas expansion (for more detail, see \cite{Fro2010}). In
other words, numerical calculation of all matrix elements with
semi-exponential functions is no more difficult problem, than for the
traditional Hylleraas radial functions. This also simplifies numerical
computation of the bound state properties (i.e. expectation values) in the
semi-exponential basis set. Our algorithms used in calculations of all
required matrix elements is based on the Perkins formula for three-electron
integrals \cite{Per} in relative coordinates. Note also that all
calculations in this work have been performed with the use of standard
quadruple precision accuracy (30 decimal digits per computer word).

In actual atomic systems any many-electron wave function must be completely
antisymmetric upon all electron variables, i.e. upon all electron spatial
and spin variables. For three-electron atomic wave function this requirement
is written in the form ${\hat{\cal A}}_{123} \Psi(1,2,3) = - \Psi(1,2,3)$,
where $\Psi$ is given by Eq.(\ref{psi}) and $\hat{{\cal A}}_e$ is the
three-particle (= electron) antisymmetrizer ${\hat{\cal A}}_e = \hat{e} -
\hat{P}_{12} - \hat{P}_{13} - \hat{P}_{23} + \hat{P}_{123} + \hat{P}_{132}$.
Here $\hat{e}$ is the identity permutation, while $\hat{P}_{ij}$ is the
permutation of the $i$-th and $j$-th particles. Analogously, the operator
$\hat{P}_{ijk}$ is the permutation of the $i$-th, $j$-th and $k$-th
particles. In actual computations antisymmetrization of the total wave
function is reduced to the proper antisymmetrization of corresponding matrix
elements (for more detail, see, e.g., \cite{Fro2010}). Each of these matrix
elements is written in the form $\langle \Psi \mid \hat{O} \mid \Psi
\rangle$, where $\hat{O}$ is an arbitrary spin-independent quantum operator
which is truly symmetric upon all interparticle permutations. The wave
function $\Psi$, Eq.(\ref{psi}), contains the two different radial parts
$\psi$ and $\phi$. By performing the integration over all spin coordinates
from here one finds the four spatial projectors ${\cal P}_{\psi\psi}, {\cal
P}_{\psi\phi} = {\cal P}_{\phi\psi}$ and ${\cal P}_{\phi\phi}$ presented in
\cite{Fro2010}. In fact, the explicit form of the ${\cal P}_{\psi\phi}$ and
${\cal P}_{\phi\psi}$ projectors given in \cite{Fro2010} must be corrected
(there is an obvious misprint in the formulas given in \cite{Fro2010})
\begin{eqnarray}
 {\cal P}_{\psi\phi} &=& \frac12 \Bigl( \hat{P}_{13} - \hat{P}_{23} +
 \hat{P}_{123} - \hat{P}_{132} \Bigr) \\
 {\cal P}_{\phi\psi} &=& \frac12 \Bigl( \hat{P}_{13} - \hat{P}_{23} +
  \hat{P}_{123} - \hat{P}_{132} \Bigr)
\end{eqnarray}
For an arbitrary truly symmetric spin-independent operator $\hat{O}$ each
of these four projectors produces matrix elements $\langle \Psi \mid \hat{O}
\mid \Psi \rangle$ of the correct permutation symmetry (for doublet states)
between all three electrons. The explicit formulas for all matrix elements
obtained with the radial basis functions, Eq.(\ref{semexp}), and for
three-electron integrals needed in calculations can be found in
\cite{Fro2010}.

\section{Calculations}

Let us apply the semi-exponential variational expansion, Eq.(\ref{semexp}),
to numerical calculations of the ground $1^2S$-state in the three-electron
${}^{\infty}$Li atom. In this study we consider the two variational wave
functions: (a) the wave function which contains 28 radial basis functions,
Eq.(\ref{semexp}), and (b) the wave function which includes 40 radial basis
functions, Eq.(\ref{semexp}). The results (in atomic units) obtained with
these two trial wave functions can be found in Table I. Tables II and III
contain the corresponding radial basis functions, Eq.(\ref{semexp}), i.e.
the powers $n_1(k), n_2(k), n_3(k), m_1(k), m_2(k), m_3(k)$ of six radial
variables $r_{12}, r_{13}, r_{23}, r_{14}, r_{24}, r_{34}$ and optimized
non-linear parameters $\alpha_k, \beta_k, \gamma_k$. As follows from Table I
our variational energies obtained for the ground $1^2S$-state in the
${}^{\infty}$Li atom with the use of semi-exponential variational expansion,
Eq.(\ref{psi}), are substantially lower than the corresponding energies
determined for this state with the same Hylleraas wave function \cite{Lars}.
Note also that the non-linear parameters used in our method (in
Eq.(\ref{semexp})) are constantly varied. Therefore, it is hard to say that
the total energies obtained in some calculations are `final'. Formally,
based on the known convergence rate(s) for our data and by using a few
extrapolation procedures we can approximately evaluate the limits to which
our variational energies will converge, if we could perform an infinite
number of variations for the non-linear parameters in Eq.(\ref{semexp}).
Such limits for the total energies are shown in the fourth column of Table
I. These values indicate that, e.g., our 40-term variational wave function
can produce, in principle, very accurate variational energies, if the
optimization of non-linear parameters in Eq.(\ref{semexp}) will continue.

In \cite{Lars} Larsson proposed a simple (but useful!) trick which allows
one to increase the overall accuracy of the trial (doublet) wave function.
Later this trick was called `doubling' of the wave function and it was used
practically in all calculations of the bound doublet states in
three-electron atomic systems. The idea of doubling is simple and
transparent. If we already know the radial function constructed for one spin
configuration, e.g., for $\chi_1 = \alpha \beta \alpha - \beta \alpha
\alpha$ from Eq.(\ref{psi}), then we can use exactly the same radial basis
function for another spin configuration $\chi_2 = 2 \alpha \alpha \beta -
\beta \alpha \alpha - \alpha \beta \alpha$. Formally, it doubles the total
number of basis functions in the trial wave function. According to the
variational principle the total variational energy can only decrease during
such a procedure. The problem of linear dependence of basis functions is
avoided in this procedure, since the two spin functions $\chi_1 = \alpha
\beta \alpha - \beta \alpha \alpha$ and $\chi_2 = 2 \alpha \alpha \beta -
\beta \alpha \alpha - \alpha \beta \alpha$ are independent of each other. In
fact, for Hylleraas variational expansion the `doubling' does not work
properly, since there are obvious linear dependencies between different
radial basis functions in those cases when some non-linear parameters
coincide with each other (for more detail, see \cite{Lars}). In
semi-exponential variational expansion Eq.(\ref{semexp}) all optimized
non-linear parameters are independent of each other. Therefore, the
coincidence of the pre-exponential factors in Eq.(\ref{semexp}) is not
crucial and does not mean that such basis functions are linearly dependent.
This drastically simplifies the actual `doubling' for Eq.(\ref{semexp}). The
energies obtained with the use of `doubling' of our variational wave
functions can be found in the fifth column of Table I. It is clear that the
non-linear parameters (or parameters in the exponents in Eq.(\ref{semexp}))
from the second part of the total wave function are not optimal, i.e. for
all terms which contain basis functions with numbers $i \ge 41$ the
non-linear parameters are not optimal. These 120 (= 40 $\times 3$)
non-linear parameters can be re-optimized and this drastically improves the
overall quality of the total wave function. For instance, approximate
re-optimization of the last 40 non-linear parameters in the wave function
gives the ground state energy -7.4780583419 $a.u.$, which is much better
than the `doubling' energy (-7.4780581691 $a.u.$) from Table I. Note also
that our value of the total energy of the ground state in the Li atom is
better than the total energy obtained in \cite{KingSoup} with the use of
352 basis functions.

As follows from Table I the doubling is not an effective approach for our
trial wave functions with the carefully optimized non-linear parameters.
However, we can modify the original idea of doubling into something new
which is substantially more effective in actual computations. To illustrate
one of such modifications, let us assume that we have constructed a 40-term
variational wave function, Eq.(\ref{semexp}) which contains 40 $\times$ 3 =
120 carefully optimized non-linear parameters $\alpha_1, \beta_1, \gamma_1,
\ldots, \alpha_{40}, \beta_{40}, \gamma_{40}$. At the second step of our
procedure we can add forty additional basis functions with the same
pre-exponential factors $r^{n_1(k)}_{23} r^{n_2(k)}_{13} r^{n_3(k)}_{12}
r^{m_1(k)}_{14} r^{m_2(k)}_{24} r^{m_3(k)}_{34}$, but slightly different
exponents in Eq.(\ref{semexp}). In reality, these new exponents have been
chosen quasi-randomly from three different intervals, e.g.,
\begin{eqnarray}
 \alpha_{i + 40} = \alpha_{i} + 0.0057 \cdot \Bigl<\Bigl< \frac{i (i + 1)
 \sqrt{2}}{2} \Bigr>\Bigr> \nonumber \\
 \beta_{i + 40} = \beta_{i} + 0.0063 \cdot \Bigl<\Bigl< \frac{i (i + 1)
 \sqrt{3}}{2} \Bigr>\Bigr> \nonumber \\
 \gamma_{i + 40} = \gamma_{i} + 0.0049 \cdot \Bigl<\Bigl< \frac{i (i + 1)
 \sqrt{5}}{2} \Bigr>\Bigr> \nonumber
\end{eqnarray}
where $i$ = 1, 2, $\ldots$, 40 and $\Bigl<\Bigl< x \Bigr>\Bigr>$ designates
the fractional part of the real number $x$. Small deviations of these new
exponents from the known `optimal' values (i.e. from $\alpha_i, \beta_i,
\gamma_i$, where $1 \le i \le 40$) produce the extended wave function of
`almost optimal' quality. On the other hand, even these small differences
between exponents allows one to avoid a linear dependence between basis
vectors in Eq.(\ref{semexp}). Obviously, this procedure can be repeated a
number of times. This allows one to construct very accurate trial wave
functions which contain not only 80, but 400, 800 and even 2000 basis
functions with almost `optimal' non-linear parameters.

It is very interesting to perform variational calculations of the ground
state of the ${}^{\infty}$Li atom with the use of 60-term wave function
constructed from the analogous 60-term wave Larsson's wave function
\cite{Lars}. The variational total energy obtained in \cite{Lars} with that
wave function and one spin function $\chi_1$ was -7.4780103597 $a.u.$ Our
60-term trial wave function with one spin function $\chi_1$ constructed in
\cite{Fro2010} from  the same radial basis functions corresponds to the
substantially lower total energy $E$ = -7.478057561 $a.u.$ The current total
energy is -7.4780597045 $a.u.$ (only one spin function $\chi_1$ is used in
our calculations). Note that our current total energy rapidly decreases with
almost constant rate $\approx 0.5 \cdot 10^{-7}$ $a.u.$ per optimization
cycle, i.e. per one variation of all 180 (= 60 $\times$ 3) non-linear
parameters in the trial wave function. This total energy is better than the
values obtained in \cite{King2} with the use of 503 Hylleraas basis
functions (selected radial configurations were used in such calculations).
Moreover, this total energy is slightly better than the value obtained in
\cite{Thak1}.

The doubling of our 60-term wave function produces the total energy
-7.4780597761 $a.u.$ This energy is close to the known `exact' answer
-7.478060323904 $a.u.$ \cite{2006}. We expect that after an infinite number
of variations of the non-linear parameters the total energy of the
${}^{\infty}$Li atom obtained with our 60-term radial function and one spin
function will converge to the value -7.4780603(3) $a.u.$ which is very close
to the actual ground state energy. It will be an outstanding result to
obtain the value lower than -7.4780602 $a.u.$ for the total energy of the
ground $1^2S-$state in the ${}^{\infty}$Li atom by using only 60-term
variational wave function.

\section{Discussions and Conclusion}

The semi-exponential variational expansion \cite{Fro2010} is applied for
bound state calculations of three-electron atomic systems. It is shown that
this variational expansion allows one to construct compact and accurate
variational wave functions for three-electron atomic systems. Currently, the
use of semi-exponential radial basis functions is the best way to produce
compact and accurate wave functions for three-electron atomic systems. The
total energies obtained in this study for the ground $1^2S$-state of the
${}^{\infty}$Li atom are more accurate than our earlier results from
\cite{Fro2010} and substantially more accurate than the original Larsson's
wave function \cite{Lars} with the same number of terms.

The results indicate clearly that our semi-exponential variational
expansion, Eq.(\ref{semexp}), has a substantially greater potential for
variational bound state calculations in three-electron atomic systems than
we have anticipated originally \cite{Fro2010}. Currently, we continue the
process of numerical optimization of the non-linear parameters in our trial
wave functions constructed with the use of semi-exponential variational
expansions. Note that the choice of Larsson's wave function(s) as the first
approximation to the semi-exponential variational expansion is not crucial
for our method. Many other choices are also possible. For instance, our next
step will be re-optimization of the non-linear parameters in the 352-term
wave function used in \cite{KingSoup}. It will take some time, but with such
a wave function we hope to produce the total energy which is close to the
known `exact' energy for the ground state of the ${}^{\infty}$Li atom (E =
-7.478060323904 $a.u.$).

When our research of the semi-exponential variational expansion started we
could not expect such impressive results. Currently, the variational wave
functions constructed with the use of semi-exponential variational expansion
are of great interest for many scientific problems which include the Li atom
and other atomic and quasi-atomic three-electron systems. In particular, we
have made numerous improvements in our original computer code
\cite{Fro2010}. Our next goal is to generalize the semi-exponential basis
to the four-electron atomic problems. The semi-exponential variational
expansion for four-electron atomic systems can be used to obtain compact and
accurate wave functions of the ground singlet $1^1S-$state and triplet
$2^3S-$states of the Be atom and Be-like ions. It is clear that our method
can also be used for rotationally excited states in atomic systems, i.e. for
states with $L \ge 1$, where $L$ is the electron angular momentum.

In conclusion, it should be mentioned that numerous attempts to improve the
overall quality of radial basis functions for three-electron atomic systems
(in general, for four-body Coulomb systems) started almost 20 years ago
(see, e.g., \cite{Fro94}, \cite{HFS03}, \cite{Reb}). In part, it was a
reaction to low efficiency of the traditional Hylleraas variational
expansion for such systems. Indeed, the total number of terms in modern
versions of Hylleraas variational expansions already exceeded 13,000 and
still growing. For the Li atom (ground state) such wave functions allows one
to determine 12 - 13 stable decimal digits in the total energy. Comparison
of a large number of numerical results for the ground state of the lithium
atom can be found in Table II from \cite{SimHar}. The new Hylleraas wave
function with 65,000 terms will allow one to obtain $\approx$ 15 stable
decimal digits in the total energy. It is clear that such a method cannot be
considered as a reasonable and appropriate approach for accurate variational
computations of the bound states in four-body systems.

An alternative approach for construction of highly accurate four-body wave
functions was proposed in \cite{Fro94}, \cite{HFS03}, \cite{Reb}. It is
based on the closed analytical formulas derived in \cite{FromHill} for the
basic four-body integral. As it follows from actual calculations this method
allows one to obtain very accurate bound state energies essentially for all
Coulomb four-body systems, including ${}^{\infty}$Li atom ($E$ =
-7.47806025114 $a.u.$, 50 radial basis functions (exponents) with the spin
function $\chi_1$ plus 19 radial basis functions with the spin function
$\chi_2$ (their definitions are given above), bi-positronium Ps$_2$ ($E$ =
-0.51600377267 $a.u.$, 50 exponential basis functions) and ${}^{\infty}$HPs
($E$ = -0.78914861151 $a.u.$, 50 exponential basis functions). Very good
values for the total (bound) state energies have also been obtained for the
$pp\mu\mu$ and $dd\mu\mu$ four-body bi-muon systems. These total energies
can be improved even further by using more careful optimization of the
non-linear parameters in the trial wave functions. For accurate computations
of the H$_2$ molecule and other similar (molecular) systems the method
\cite{HFS03} must be modified to include the complex values for some
non-linear parameters (as was done in \cite{Reb}).

Note, however, that relations between these variational expansions based on
the formula from \cite{FromHill} and traditional Hylleraas variational
expansion are extremely complicated. Further analysis shows that the source
of these difficulties is directly related with the fundamental properties of
the four-body perimetric coordinates \cite{Fro06}, rather than with the lack
of `good representations' for the formula Eqs.(2.1) - (2.9) from
\cite{FromHill}. As follows from the properties of the four-body perimetric
coordinates \cite{Fro06} it is very hard to represent any of these compact
and accurate `exponential' wave functions in terms of the Hylleraas basis
functions. The overall complexity of problems arising here is comparable
with the difficulties which can be found in the original problem. Many
advantages of the `exponential' wave functions, e.g., their very accuracy,
can be lost during this procedure. In many cases, it is simpler to
re-calculate the corresponding energies from the very beginning by using the
traditional Hylleraas variational expansion. This is the main reason why we
do not want to discuss here some recent developments in the four-body
exponential and related variational expansions (see, e.g.,
\cite{Harris2009}).

Our approach allows one to produce compact and accurate wave functions which
are easily related with the Hylleraas basis set of radial functions. This
means that all such compact and accurate wave functions can directly be
used in computations of various bound state properties and transition
probabilities. The second advantage of the semi-exponential variational
expansion follows from relatively simple formulas for all matrix elements
needed in highly accurate computations and optimization processes.

\newpage
%
% TABLE I
%
   \begin{table}
    \caption{The total energies $E$ (in atomic units) of the $1^2S(L =
             0)-$state in the ${}^{\infty}$Li atom. $N$ designates the
             number of basis functions used.}
      \begin{center}
      \begin{tabular}{lllll}
      \hline\hline
 $N$ & $E$(Ref.[4]) & $E$(Eq.(3)) & $E^{a}$(Eq.(3)) & $E$(Eq.(3); doubling) \\
     \hline
 28 & -7.477885105 & -7.4780363801 & -7.4780368(3) & -7.4780365786 \\

 40 & -7.477944869 & -7.4780579457 & -7.4780595(5) & -7.4780580161 \\
  \hline\hline
  \end{tabular}
  \end{center}
  \end{table}
%
% TABLE II
%
  \begin{table}[tbp]
   \caption{An example of the trial wave function constructed with the use
            of $N = 28$ semi-exponential radial basis functions. This wave
            function produces the total energy $E$ = -7.4780363801 $a.u.$
            for the ground $1^2S-$state of the ${}^{\infty}$Li atom. Only
            one electron spin-function $\chi_1 = \alpha \beta \alpha -
            \beta \alpha \alpha$ was used in these calculations.}
     \begin{center}
     \scalebox{0.72}{%
     \begin{tabular}{cccccccccc}
      \hline\hline
 $N$ & $n_1$ & $n_2$ & $n_3$ & $m_1$ & $m_2$ & $m_3$ & $\alpha$ & $\beta$ & $\gamma$ \\
     \hline
 1  &  0 &  0 &   0 &   0 &   0 &   1 & 0.340570905705403E+01 & 0.293295186982955E+01 & 0.771546188231103E+00 \\

 2  &  0 &  0 &   0 &   1 &   0 &   1 & 0.182145794896563E+01 & 0.329011437023099E+01 & 0.316360558089102E+01 \\

 3  &  0 &  0 &   0 &   1 &   1 &   1 & 0.276859627151255E+01 & 0.297627929771076E+01 & 0.668289237018917E+00 \\

 4  &  0 &  0 &   0 &   2 &   0 &   1 & 0.286729169596989E+01 & 0.300017942427107E+01 & 0.637008666805219E+00 \\

 5  &  0 &  0 &   1 &   0 &   0 &   1 & 0.278208405003856E+01 & 0.275168552095737E+01 & 0.645451709771134E+00 \\

 6  &  0 &  0 &   2 &   0 &   0 &   1 & 0.373836260739635E+01 & 0.339167728695617E+01 & 0.658338285572628E+00 \\

 7  &  0 &  0 &   0 &   0 &   0 &   0 & 0.337994212648889E+01 & 0.324225523017272E+01 & 0.108564332122329E+01 \\

 8  &  1 &  0 &   0 &   0 &   0 &   0 & 0.167110811817316E+01 & 0.338626357605054E+01 & 0.867142603936635E+00 \\

 9  &  0 &  0 &   0 &   0 &   0 &   2 & 0.306529447275211E+01 & 0.299446682705661E+01 & 0.688809932201360E+00 \\

 10 &  1 &  0 &   0 &   0 &   1 &   0 & 0.263305048764256E+01 & 0.242503640146648E+01 & 0.776895925198369E+00 \\

 11 &  0 &  0 &   3 &   0 &   0 &   1 & 0.521025323497647E+01 & 0.361243051429800E+01 & 0.660062464430887E+00 \\

 12 &  1 &  0 &   0 &   0 &   0 &   1 & 0.306893947287392E+01 & 0.208144332232383E+01 & 0.123105191974125E+01 \\

 13 &  0 &  0 &   0 &   0 &   0 &   3 & 0.186312473874842E+01 & 0.289718514568496E+01 & 0.996173732965274E+00 \\

 14 &  0 &  0 &   1 &   1 &   0 &   1 & 0.454831261352940E+01 & 0.219828523532609E+01 & 0.158185067718071E+01 \\

 15 &  0 &  0 &   0 &   3 &   0 &   1 & 0.295559769377747E+01 & 0.327296616683230E+01 & 0.672095727321017E+00 \\

 16 &  0 &  0 &   4 &   0 &   0 &   1 & 0.113723992478698E+02 & 0.508008324231496E+01 & 0.636162427376743E+00 \\

 17 &  0 &  0 &   0 &   2 &   2 &   1 & 0.368595304619283E+01 & 0.278359204737422E+01 & 0.620935860917418E+00 \\

 18 &  0 &  0 &   1 &   1 &   1 &   1 & 0.357466607750891E+01 & 0.343630387369804E+01 & 0.698563354607838E+00 \\

 19 &  0 &  0 &   1 &   2 &   0 &   1 & 0.365095130448352E+01 & 0.317945977760753E+01 & 0.652269908234656E+00 \\

 20 &  0 &  0 &   1 &   3 &   0 &   1 & 0.290005170216145E+01 & 0.289825235917765E+01 & 0.685753232846547E+00 \\

 21 &  2 &  0 &   0 &   0 &   0 &   0 & 0.280340788922781E+01 & 0.225117021991187E+01 & 0.990926897297406E+00 \\

 22 &  1 &  1 &   0 &   0 &   0 &   0 & 0.275272575017714E+01 & 0.276113990248958E+01 & 0.802457922378512E+00 \\

 23 &  1 &  0 &   0 &   0 &   2 &   0 & 0.606571100003296E+01 & 0.326870035708320E+01 & 0.453868977046091E+00 \\

 24 &  1 &  0 &   0 &   1 &   1 &   0 & 0.293743657648302E+01 & 0.226578275529031E+01 & 0.669408905271258E+00 \\

 25 &  0 &  0 &   0 &   0 &   0 &   4 & 0.502952034298549E+01 & 0.276772288773810E+01 & 0.950599883558449E+00 \\

 26 &  0 &  0 &   1 &   0 &   0 &   0 & 0.868992405342042E+01 & 0.374899612568756E+01 & 0.516228416453001E+00 \\

 27 &  0 &  0 &   1 &   0 &   0 &   2 & 0.433237156102395E+01 & 0.278286578531691E+01 & 0.112727477282094E+01 \\

 28 &  0 &  0 &   0 &   1 &   0 &   0 & 0.221163211952812E+01 & 0.498938978770980E+01 & 0.647628325318278E+00 \\
  \hline\hline
  \end{tabular}}
  \end{center}
  \end{table}
%
% TABLE III
%
  \begin{table}[tbp]
   \caption{An example of the trial wave function constructed with the use
            of $N = 40$ semi-exponential radial basis functions. This wave
            function produces the total energy $E$ = -7.47805542591 $a.u.$
            for the ground $1^2S-$state of the ${}^{\infty}$Li atom. Only
            one electron spin-function $\chi_1 = \alpha \beta \alpha -
            \beta \alpha \alpha$ was used in these calculations.}
     \begin{center}
     \scalebox{0.62}{%
     \begin{tabular}{cccccccccc}
      \hline\hline
 $N$ & $n_1$ & $n_2$ & $n_3$ & $m_1$ & $m_2$ & $m_3$ & $\alpha$ & $\beta$ & $\gamma$ \\
     \hline
 1  &  0 &  0 &   0 &   0 &   0 &   1 & 0.310888325274582E+01 & 0.290883960347876E+01 & 0.845418768108028E+00 \\

 2  &  0 &  0 &   0 &   1 &   0 &   1 & 0.196342255489068E+01 & 0.347479833068431E+01 & 0.163523383561211E+01 \\

 3  &  0 &  0 &   0 &   1 &   1 &   1 & 0.278598083293758E+01 & 0.282092658543858E+01 & 0.679130247211775E+00 \\

 4  &  0 &  0 &   0 &   2 &   0 &   1 & 0.287878476756554E+01 & 0.284028201233856E+01 & 0.652121055380736E+00 \\

 5  &  0 &  0 &   1 &   0 &   0 &   1 & 0.296290423881450E+01 & 0.281951549636015E+01 & 0.660626319649503E+00 \\

 6  &  0 &  0 &   2 &   0 &   0 &   1 & 0.356029841316233E+01 & 0.298528920333113E+01 & 0.672210253693381E+00 \\

 7  &  0 &  0 &   0 &   0 &   0 &   0 & 0.281820369226934E+01 & 0.307242548167154E+01 & 0.614180061849861E+00 \\

 8  &  1 &  0 &   0 &   0 &   0 &   0 & 0.129540716567888E+01 & 0.388006326151676E+01 & 0.588477295803696E+00 \\

 9  &  0 &  0 &   0 &   0 &   0 &   2 & 0.312600561158126E+01 & 0.298186247192942E+01 & 0.711199626874967E+00 \\

 10 &  1 &  0 &   0 &   0 &   1 &   0 & 0.309479768715561E+01 & 0.205912634163624E+01 & 0.692766055505988E+00 \\

 11 &  0 &  0 &   3 &   0 &   0 &   1 & 0.421833159497161E+01 & 0.323537515175651E+01 & 0.683447854994486E+00 \\

 12 &  1 &  0 &   0 &   0 &   0 &   1 & 0.317935937791374E+01 & 0.246672317399557E+01 & 0.128821985777175E+01 \\

 13 &  0 &  0 &   0 &   0 &   0 &   3 & 0.206773619817191E+01 & 0.258067622963825E+01 & 0.102117374065975E+01 \\

 14 &  0 &  0 &   1 &   1 &   0 &   1 & 0.407929602782570E+01 & 0.308726834443555E+01 & 0.145857028906218E+01 \\

 15 &  0 &  0 &   0 &   3 &   0 &   1 & 0.290479112486719E+01 & 0.342131088097724E+01 & 0.654819087759473E+00 \\

 16 &  0 &  0 &   4 &   0 &   0 &   1 & 0.759200717376427E+01 & 0.403600107010476E+01 & 0.654094476238819E+00 \\

 17 &  0 &  0 &   0 &   2 &   2 &   1 & 0.342043341882141E+01 & 0.300593652535184E+01 & 0.758980390711669E+00 \\

 18 &  0 &  0 &   1 &   1 &   1 &   1 & 0.337729059232100E+01 & 0.336098394507054E+01 & 0.689553038540639E+00 \\

 19 &  0 &  0 &   1 &   2 &   0 &   1 & 0.376783902187767E+01 & 0.308351505579080E+01 & 0.693591996711379E+00 \\

 20 &  0 &  0 &   1 &   3 &   0 &   1 & 0.341103485088192E+01 & 0.390874343351408E+01 & 0.353215493102457E+01 \\

 21 &  2 &  0 &   0 &   0 &   0 &   0 & 0.187375370334190E+01 & 0.292204564107203E+01 & 0.134200544832659E+01 \\

 22 &  1 &  1 &   0 &   0 &   0 &   0 & 0.203602651426972E+01 & 0.251346948989043E+01 & 0.139512883704298E+01 \\

 23 &  1 &  0 &   0 &   0 &   2 &   0 & 0.307587342243077E+01 & 0.289804349780859E+01 & 0.799088792948691E+00 \\

 24 &  1 &  0 &   0 &   1 &   1 &   0 & 0.237469500100334E+01 & 0.411142232808047E+01 & 0.113515551805206E+01 \\

 25 &  0 &  0 &   0 &   0 &   0 &   4 & 0.306630970495110E+01 & 0.358372314305121E+01 & 0.865024537355183E+00 \\

 26 &  0 &  0 &   1 &   0 &   0 &   0 & 0.652799588882586E+01 & 0.391736310812260E+01 & 0.504690239191085E+00 \\

 27 &  0 &  0 &   1 &   0 &   0 &   2 & 0.523338852365502E+01 & 0.400156315581741E+01 & 0.119987716494855E+01 \\

 28 &  0 &  0 &   0 &   1 &   0 &   0 & 0.303300703612990E+01 & 0.269004008151858E+01 & 0.132439178382705E+01 \\

 29 &  0 &  0 &   0 &   1 &   0 &   2 & 0.347252909237306E+01 & 0.351142344973601E+01 & 0.841746594395407E+00 \\

 30 &  1 &  0 &   1 &   0 &   0 &   0 & 0.246738217080682E+01 & 0.268143054665150E+01 & 0.930487658749999E+00 \\

 31 &  2 &  0 &   0 &   0 &   1 &   0 & 0.313058760564443E+01 & 0.295867936877980E+01 & 0.903526045564038E+00 \\

 32 &  1 &  0 &   0 &   0 &   1 &   1 & 0.324981907924628E+01 & 0.322474651197190E+01 & 0.896978954638563E+00 \\

 33 &  3 &  0 &   0 &   0 &   0 &   0 & 0.333729310041323E+01 & 0.277869134235698E+01 & 0.100059195659526E+01 \\

 34 &  2 &  0 &   0 &   0 &   0 &   1 & 0.345708067187951E+01 & 0.277492748358036E+01 & 0.104744023031760E+01 \\

 35 &  0 &  0 &   5 &   0 &   0 &   1 & 0.621080396345153E+01 & 0.394717905188376E+01 & 0.771357719659273E+00 \\

 36 &  0 &  0 &   0 &   4 &   0 &   1 & 0.617132034807857E+01 & 0.257389911129776E+01 & 0.405076099224201E+01 \\

 37 &  0 &  0 &   1 &   4 &   0 &   1 & 0.384658211093920E+01 & 0.426365221909316E+01 & 0.963468025067806E+00 \\

 38 &  0 &  0 &   0 &   5 &   0 &   1 & 0.358540307325767E+01 & 0.426395800491431E+01 & 0.120557474663118E+01 \\

 39 &  0 &  0 &   2 &   1 &   0 &   1 & 0.290948705798909E+01 & 0.342780417090815E+01 & 0.803444693562464E+00 \\

 40 &  0 &  0 &   2 &   2 &   0 &   1 & 0.348532012166237E+01 & 0.473382479410345E+01 & 0.871338172104194E+00 \\
  \hline\hline
  \end{tabular}}
  \end{center}
  \end{table}
\end{document}